\documentclass[preprintnumbers,showpacs,amsmath,amssymb,floatfix,9pt,prd,onecolumn,
superscriptaddress,nofootinbib]{revtex4}
\usepackage{latexsym}
\usepackage{epsfig}
\usepackage{amssymb}

\begin{document}

\title{Nonommutative wormholes in $f(R)$ gravity}

\author{ Mubasher Jamil}
\email{mjamil@sns.nust.edu.pk} \affiliation {School of Natural
Sciences, National University of Sciences and Technology (NUST),
H-12, Islamabad, Pakistan}

\author{ Farook Rahaman}
\email{rahaman@iucaa.ernet.in} \affiliation{Department of
Mathematics, Jadavpur University, Kolkata 700 032, West Bengal,
India}
\author{ Ratbay Myrzakulov}
\email{rmyrzakulov@gmail.com} \affiliation{Eurasian International
Center for Theoretical Physics,
 Eurasian National University, Astana 010008, Kazakhstan}

\author{ P.K.F. Kuhfittig}
\email{kuhfitti@msoel.edu} \affiliation{ Department of
Mathematics,Milwaukee School of Engineering,\\  \small Milwaukee,
Wisconsin 53202-3109,USA}

\author{Nasr Ahmed}
\email{abualansar@gmail.com} \affiliation{Astronomy Department,
National Research Institute of Astronomy and Geophysics, Helwan,
Egypt}

\author{ Umar F Mondal}
\email{umarfarooquemondal@ymail.com} \affiliation{Department of
Mathematics Behala College, Parnasree, Kolkata 700060, India}

\begin{abstract}
{\bf Abstract:}  This paper discusses several new exact solutions
of static wormholes in $f(R)$ gravity with a
noncommutative-geometry background, which replaces point-like
structures by smeared objects.  In the first part of the paper
we assume the power-law form $f(R)=aR^n$ and discuss several
solutions corresponding to different values of the exponent.
The second part of the paper assumes a particular form of the
shape function that also yields a viable solution.  This
investigation generalizes some of our previous work in $f(R)$
gravity, as well as in noncommutative geometry.

{\bf Keywords:} wormholes; modified gravity; energy conditions
\end{abstract}

\pacs{04.50.-h, 04.50.Kd, 04.20.Jb}

\maketitle

\section{Introduction}

\noindent
Wormholes are topological tunnel-like structures connecting
different regions of spacetime. Once believed to be
submicroscopic, it was shown in 1988 \cite{mo} that wormholes
could be large enough for humanoid travelers and even permit
time travel.  Since then an enormous number of new wormhole
solutions and their astrophysical implications in various gravity
theories have been proposed. (See, for instance, \cite{ji} and
references therein.) Most recently, static wormholes have been
explored in generalized teleparallel gravity as well
\cite{jamil1}.

An interesting and important development of string theory is  the
realization that coordinates determining the geometry may become
noncommutative operators in a $D$-brane \cite{1,2}. This results
in a fundamental discretization of spacetime due to the commutator
$[\textbf{x}^\mu, \textbf{x}^\nu]=i\theta^{\mu\nu}$, where
$\theta^{\mu\nu}$ is an antisymmetric matrix. There is an
interesting similarity to the uncertainty principle in which the
Planck constant $\hbar$ discretizes phase space \cite{3}. This
noncommutative geometry is an intrinsic property of spacetime that
does not depend on particular features such as curvature. Moreover,
it was pointed out in Ref. \cite{4} that noncommutativity replaces
pointlike structures by smeared objects, thereby eliminating the
divergences that normally appear in general relativity. This
smearing can be modeled by the use of the Gaussian distribution of
minimal length $\sqrt{\theta}$ instead of the Dirac delta function.
So the energy density of the static and spherically symmetric
smeared and particlelike gravitational source has the form \cite{5}
\begin{equation}
  \rho(r)=\frac{M}{(4\pi\theta)^{3/2}}e^{-\frac{r^2}{4\theta}}.
\end{equation}
The mass $M$ could be a diffused centralized object such as a
wormhole \cite{6}. The Gaussian source has also been used by Sushkov
\cite{7} to model phantom-energy supported wormholes, as well as by
Nicolini and Spalluci \cite{8} for the purpose of modeling the
physical effects of short-distance fluctuations of noncommutative
coordinates in the study of black holes. Galactic rotation curves
inspired by a noncommutative-geometry background are discussed in
one of our recent works \cite{9}. The stability of a particular
class of thin-shell wormholes in noncommutative geometry is
analyzed in Ref. \cite{10}.

Recently Rahaman et al. \cite{farook} investigated higher-dimensional
wormholes in Einstein gravity with noncommutative geometry. It
was shown that wormhole solutions exist in the usual four, as well as
in five dimensions, but they do not exist in higher-dimensional
spacetimes. Based on that study, Kuhfittig \cite{k} showed that
wormholes in noncommutative geometry can be macroscopic. The
necessary violation of the weak energy condition is attributable to
the noncommutative-geometry background rather than to the use of
exotic matter. He concluded that if string theory is correct, then
the laws of physics allow macroscopic traversable wormholes with
zero tidal forces that are stable to linearized radial perturbations.

In this paper we derive some new exact solutions of static
wormholes in $f(R)$ gravity in a noncommutative-geometry setting.
Section III assumes the power-law form $F(R)=aR^{n-1}$ and
discusses several solutions corresponding to different values of
$n$, including the special cases $n=1$ (Einstein gravity) and
$n=2$ ($R^2$ gravity). Section IV assumes a particular form of the
shape function that also yields a viable solution.  Common to all
these solutions is the absence of tidal forces.

\section{Field equations  in $F(R)$ gravity}

\noindent
To describe a spherically symmetric wormhole spacetime, we take the
metric to be
\begin{equation}
ds^2= - e^{2\Phi(r)} dt^2+ \frac{dr^2}{1-b(r)/r}+r^2
(d\theta^2+\sin^2\theta\, d\phi^2). \label{Eq3}
\end{equation}
Here $\Phi(r)$ is a gravitational redshift function and $b(r)$ is
the shape function. The most general energy-momentum tensor is given
by \cite{lobo}
\begin{equation}
T_\nu^\mu=  ( \rho + p_r)u^{\mu}u_{\nu} - p_r g^{\mu}_{\nu}+ (p_t
-p_r )\eta^{\mu}\eta_{\nu}, \label{eq:emten}
\end{equation}
where $u^{\mu}u_{\mu} = - \eta^{\mu}\eta_{\mu} = 1 $ and $u^\mu
\eta_\mu=0.$ Here the vector $u^\mu$ is the fluid 4-velocity and
$\eta^\mu$ is a space-like vector orthogonal to $u^\mu$.

According to Ref. \cite{lobo}, the gravitational field equations in
$f(R)$ gravity can be written as
\begin{eqnarray}
\rho(r) &=& \frac{F b'}{r^2},\label{r}\\
p_r(r)&=& -\frac{F b}{r^3} +\frac{F'}{2r^2}(b'r-b)-F''
\left(1-\frac{b}{r}\right),\label{r1}\\
p_t(r)&=& -\frac{F'}{r}\left(1-\frac{b}{r}\right)
-\frac{F}{2r^3}(b'r-b).\label{r2}
\end{eqnarray}
The above equations   are the generic expressions for the matter
threading the wormhole as a function of the shape function, as well
as the specific form of $F(r)$, where $F =\frac{df}{dR}$.  Here the
prime denotes the  derivative with respect to $r$.

The curvature scalar $R$ is given by
\begin{equation}\label{Ricci}
    R(r) = 2\frac{b'}{r^2}.
   \end{equation}

\section{Wormholes for given $F(R)$ functions}
\noindent
The past several decades have seen an immense interest by theorists
in searching  for viable alternative theories of modified gravity.
The basic intent in modifying or extending general relativity was
to explain certain cosmic phenomena such as dark matter, cosmic
inflation in the early Universe, and the present cosmic accelerated
expansion \cite{jamil}. Furthermore ``it is of interest to study
gravitational theories which are diffeomorphism invariant and give
Einstein gravity in an appropriate limit, but deviate from Einstein
gravity in some way outside of the realm where gravitational effects
have commonly been observed \cite{lit}.'' Our interest is confined
to $f(R)$ gravity, where $R$ is the Ricci scalar. This gravitational
theory has attained several theoretical and observational triumphs
in recent years. (For reviews see \cite{fr} and references therein.)

We are going to assume a constant redshift function for our model,
the so-called zero-tidal force solution, i.e., $\Phi(r)=\Phi_0$
(where $\Phi_0$ is a constant) and a power-law form
\begin{equation}\label{power}
F(R) =a R^{n-1}.
\end{equation}
Here $a$ is a constant and $n$ is an integer. It is worth noting
that wormholes with power-law $F(R)$ gravity and a non-constant
redshift function have been explored in the literature \cite{lit},
but for simplicity and viability purposes, we shall restrict
ourselves to the above assumption.

In solving the field equations, we have taken the energy density
of the static and spherically symmetric smeared and particle-like
gravitational source to be of the form given in Eq. (1):
\begin{equation}\label{rho}
  \rho(r)=\frac{M}{(4\pi\theta)^{\frac{3}{2}}}e^{-\frac{r^2}{4\theta}};
\end{equation}
as already noted, the mass $M$ could be a diffused centralized object
such as a wormhole.

We obtain the shape function $b(r)$ by solving the differential
equation obtained from Eqs. (\ref{r}), (\ref{Ricci}), (\ref{power}),
and (\ref{rho}).
The result is
\begin{equation}\label{shape}
b(r) = m_0 \left[ -2rn \theta e^{-\frac{r^2}{4n\theta}} +
2n^{\frac{3}{2}} \theta^{\frac{3}{2}} \pi^{\frac{1}{2}} erf
\left \{{\frac{r}{2\sqrt{n\theta}}}\right\}+C \right],
\end{equation}
where
\[
m_0 = \Big(\frac{M}{2^{n-1} a (4\pi \theta)^{\frac{3}{2}}
}\Big)^{\frac{1}n},
\]
erf$(x)=\frac{2}{\sqrt{\pi}}\int\limits_0^{x}e^{-t^2}dt$,
the error function, and $C$ is an integration constant.
It is easily checked that $R$ becomes
\begin{equation}
R=2\Big( \frac{M}{ a 2^{n-1}(4\pi\theta)^{\frac{3}{2}}}
\Big)^{\frac{1}{n}}e^{-\frac{r^2}{4n\theta}}.
\end{equation}

In
Eq. (\ref{shape}), $n=1$ corresponds to Einstein gravity,
also obtained in Ref. \cite {farook}; the case $n=2$ is
commonly referred to as quadratic or $R^2$ gravity.

To get the exact physical characteristics, we will discuss
several models resulting from  different choices of $n$.
\\
\\
\\
\textbf{Subcase I:} $n=1$
\\
\\
By Eq. (\ref{power}), the assumption $n=1$ implies that we
are dealing with Einstein gravity with a
noncommutative-geometry background.  Here the shape function
assumes the form

\begin{equation}
b(r) = m_0 \left[ -2r  \theta e^{-\frac{r^2}{4\theta}} +2
  \theta^{\frac{3}{2}} ( \pi)^{\frac{1}{2}} erf \left
\{{\frac{r}{2\sqrt{ \theta}}}\right\}+C \right],
\end{equation}
where   $$m_0 = \frac{M}{a (4\pi \theta)^{\frac{3}{2}}}.$$
\\
This result agrees with our earlier result \cite{farook}.
\\
\\
\textbf{Subcase II:} $n=2$
\\
\\
The assumption $n=2$ corresponds to $R^2$ gravity with a
noncommutative-geometry background.  Now the shape function
 takes on the form

\begin{equation}
b(r) = m_0 \left[ -4r  \theta e^{-\frac{r^2}{8\theta}} + 4
  \theta^{\frac{3}{2}} (2\pi)^{\frac{1}{2}} erf \left
\{{\frac{r}{2\sqrt{2\theta}}}\right\} +C\right],
\end{equation}
where   $$m_0 = \Big(\frac{M}{2  a (4\pi \theta)^{\frac{3}{2}}
}\Big)^{\frac{1}{2}}.$$

The next step is to check that the shape function leads to the
required wormhole structure. Using some typical values of the
parameters, the resulting shape function is pictured in Fig. 1 (left
panel).  The middle panel shows that $b(r)/r\rightarrow0$ as
$r\rightarrow\infty$; so if the constant redshift function is joined
smoothly to a function going to zero as $r\rightarrow\infty$, the
spacetime becomes asymptotically flat. The throat of the wormhole is
located at $r=0.35$, where $G(r)=b(r)-r$ cuts the $r$-axis, shown in
Fig. 1 (right panel), also born out by Fig. 1 (left panel).  In
addition, Fig. 1 indicates that for $r > r_0$, $G(r)<0$, i.e.,
$b(r)-r<0$, which implies that $\frac{b(r)}{r}<1$ for $r>r_0$ (where
$r_0$ is the radius or size of the wormhole's throat), an essential
requirement for a shape function. Also, $G(r)$ is a decreasing
function for $r> r_0$. Since $G'(r)<0$, we have $b'(r_0)<1$, which
is the flare-out condition. It now becomes apparent that the shape
function has produced the desired wormhole structure.

It is interesting to note that on a cosmological scale, quadratic
$R^2$ gravity is considered physically viable since it is
renormalizable and has numerous astrophysical implications
\cite{ABS}.

The radial pressure component is given by
\begin{eqnarray}
p_r(r)= -\frac{2 a m_0e^{-\frac{r^2}{8\theta}}}{r^3}\left[m_0
\left\{ -4r  \theta e^{-\frac{r^2}{8\theta}} + 4
  \theta^{\frac{3}{2}} (2\pi)^{\frac{1}{2}} erf \left
({\frac{r}{2\sqrt{2\theta}}}\right) +C\right\} \right]
\nonumber\\-\frac{ a m_0e^{-\frac{r^2}{8\theta}}}{2 \theta r
}\left[ m_0 r^3e^{-\frac{r^2}{8\theta}} -m_0 \left\{ -4r  \theta
e^{-\frac{r^2}{8\theta}} + 4
  \theta^{\frac{3}{2}} (2\pi)^{\frac{1}{2}} erf \left
({\frac{r}{2\sqrt{2\theta}}}\right) +C\right\} \right] \nonumber\\
-\left[ -\frac{ a m_0e^{-\frac{r^2}{8\theta}}}{2 \theta}+\frac{a
m_0 r^2 e^{-\frac{r^2}{8\theta}}}{8 \theta^2} \right] \left[1 -
\frac{m_0}{r} \left\{ -4r \theta e^{-\frac{r^2}{8\theta}} + 4
  \theta^{\frac{3}{2}} (3\pi)^{\frac{1}{2}} erf \left
({\frac{r}{2\sqrt{2\theta}}}\right) +C\right\}\right]
\end{eqnarray}

and the transverse pressure component is
\begin{eqnarray}
p_t(r)=\left[ \frac{ a m_0e^{-\frac{r^2}{8\theta}}}{2 \theta}
\right]    \left[1 -  \frac{m_0}{r} \left\{ -4r \theta
e^{-\frac{r^2}{8\theta}} + 4
  \theta^{\frac{3}{2}} (2\pi)^{\frac{1}{2}} erf \left
({\frac{r}{2\sqrt{2\theta}}}\right) +C\right\}\right]
\nonumber\\
-\frac{ a m_0e^{-\frac{r^2}{8\theta}}}{ r^3 }\left[ m_0
r^3e^{-\frac{r^2}{8\theta}} -m_0 \left\{ -4r  \theta
e^{-\frac{r^2}{8\theta}} + 4
  \theta^{\frac{3}{2}} (2\pi)^{\frac{1}{2}} erf \left
({\frac{r}{2\sqrt{2\theta}}}\right)+C \right\} \right].
\end{eqnarray}
\begin{figure*}[thbp]
\begin{tabular}{rl}
\includegraphics[width=4.5cm]{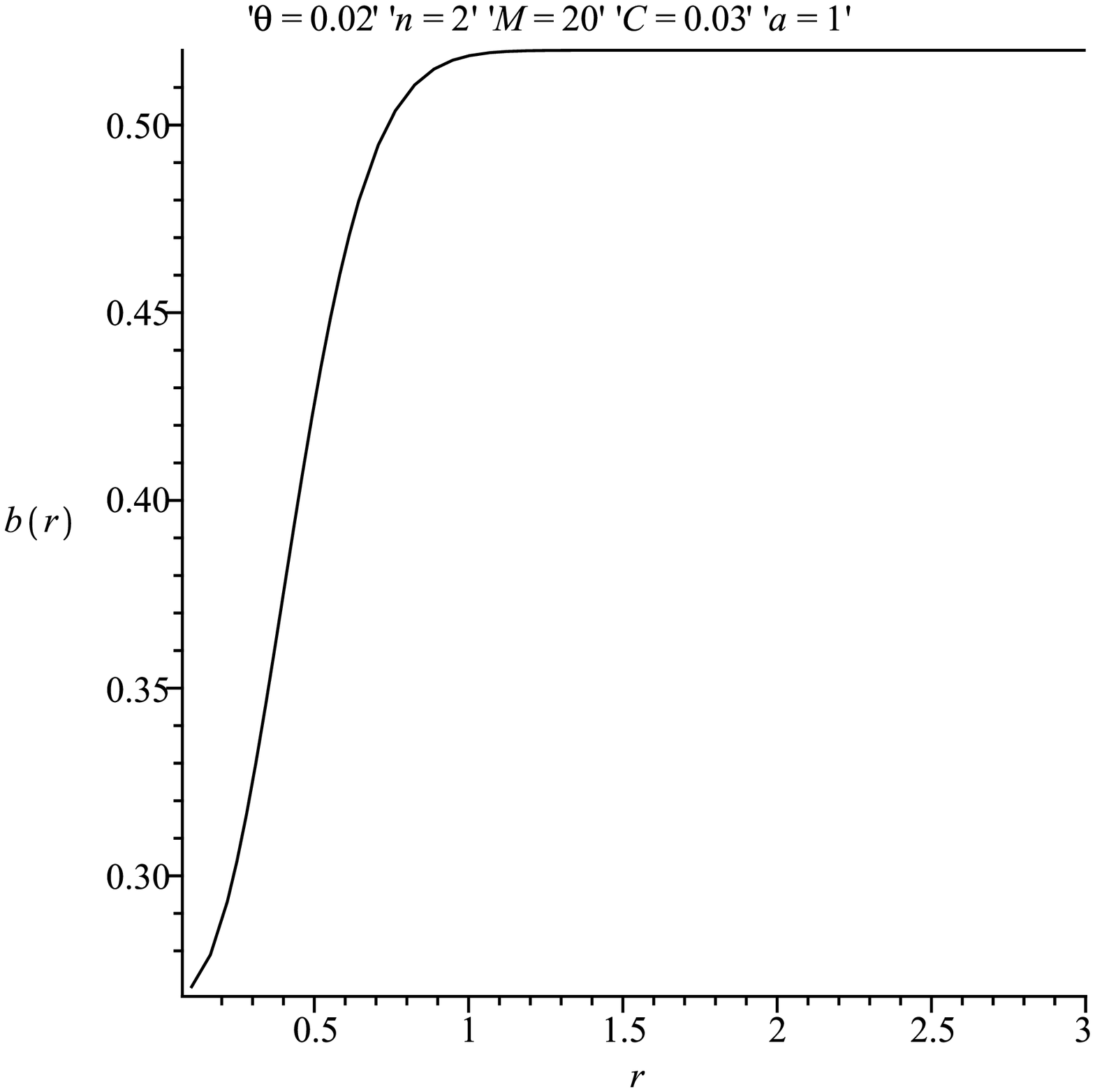}&
\includegraphics[width=4.5cm]{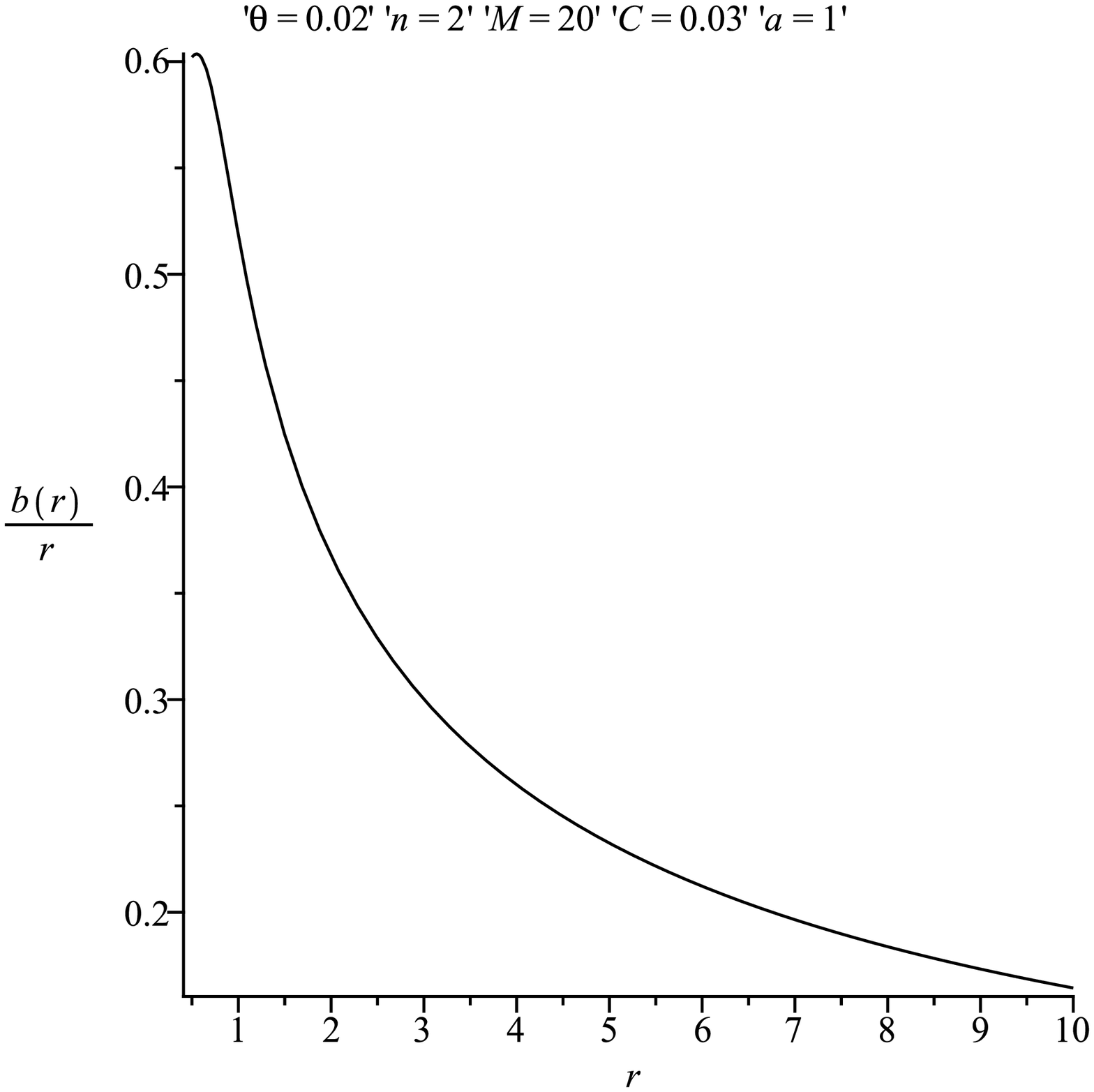}
\includegraphics[width=4.5cm]{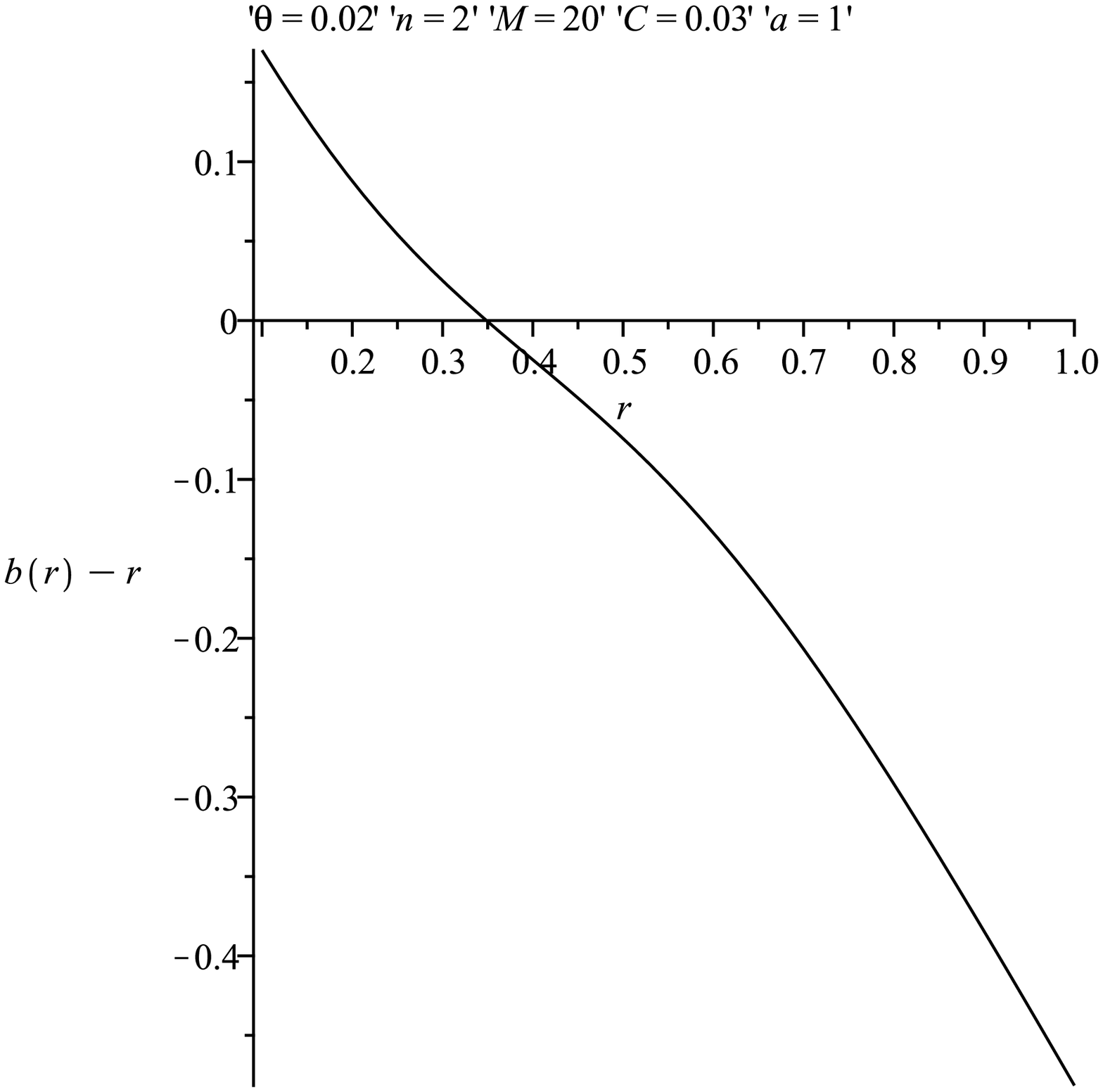} \\
\end{tabular}
\caption{ (\textit{Left})  Diagram of the shape function of the
wormhole in $R^2$ gravity for specific values of the parameters:
$\theta =0.02$, $M=20$, $a=1$, and $C=0.03$. (\textit{Middle})
Asymptotic behavior of the shape function. (\textit{Right})
The throat occurs where $G(r) = b(r)-r$ cuts the $r$-axis.}
\end{figure*}
Fig. 2 (right  panel) indicates that  the null energy condition
is violated.  Fig. 2 (left panel) shows the derivative of the
shape function with respect to $r$.

\begin{figure*}[thbp]
\begin{tabular}{rl}
\includegraphics[width=5.5cm]{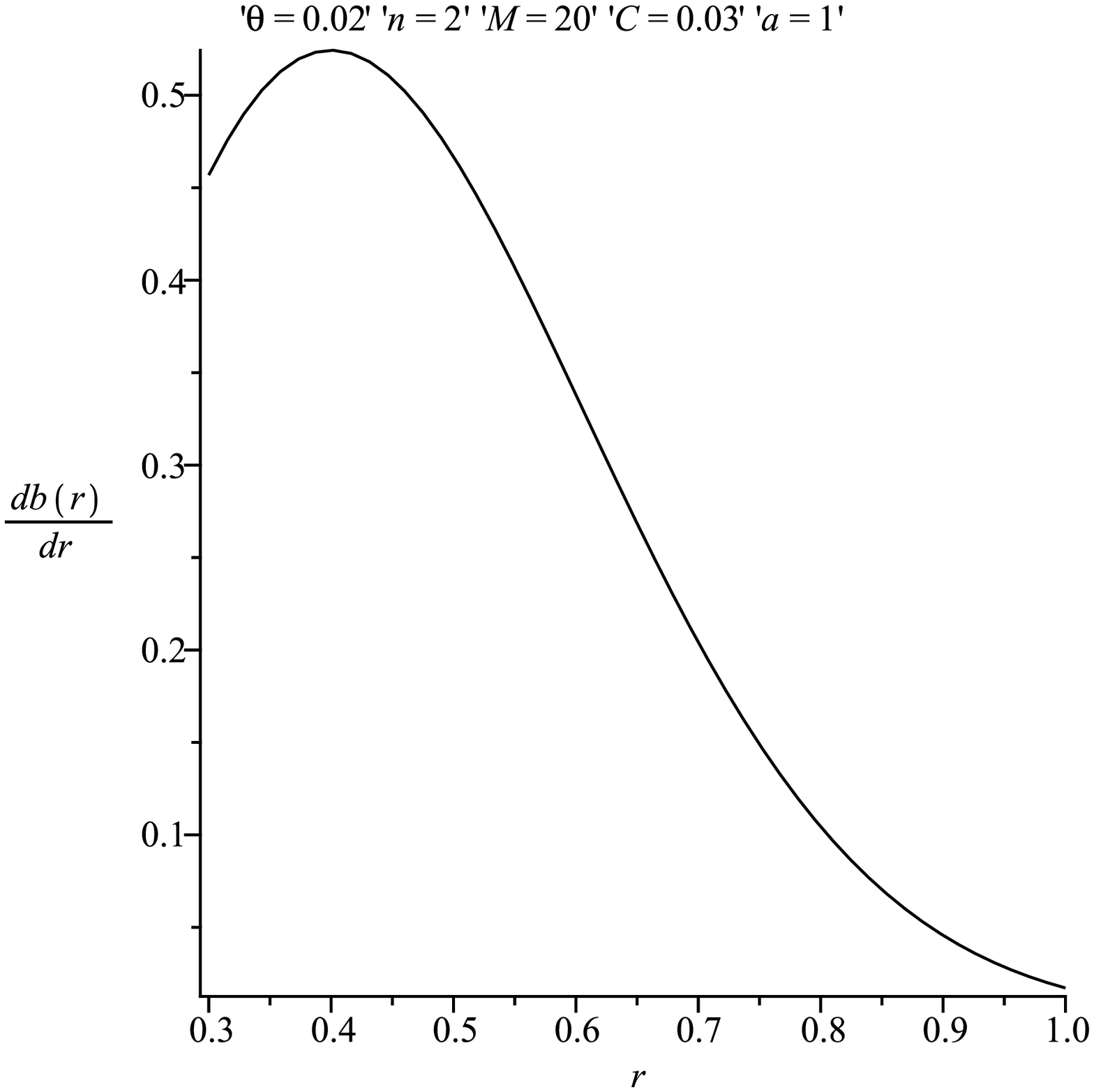}&
\includegraphics[width=5.5cm]{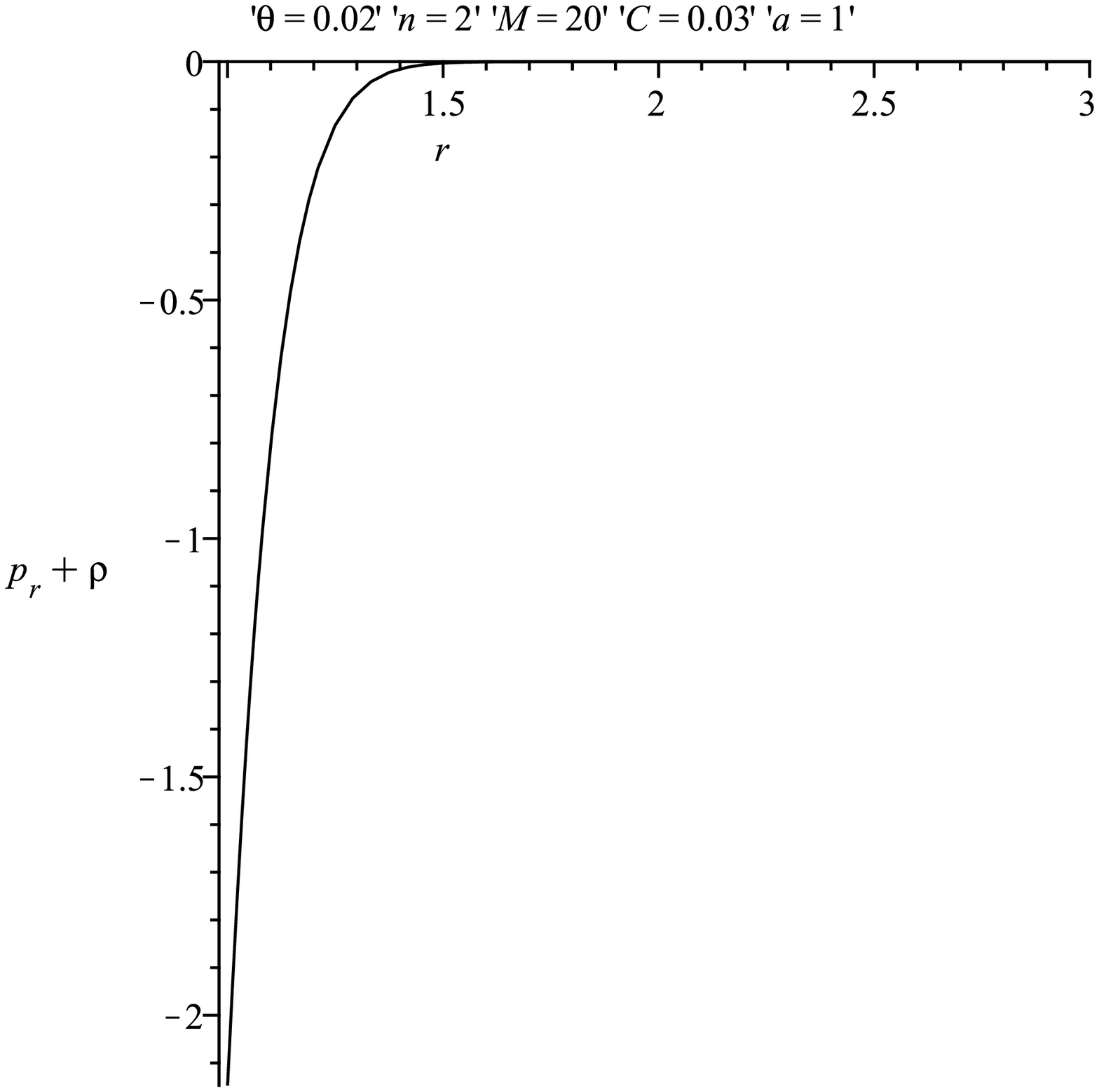} \\
\end{tabular}
\caption{ (\textit{Left})     Diagram of the derivative of
the shape function of the wormhole. (\textit{Right}) The
variation of the left-hand side of the expression for the null
energy condition with respect to $r$.}
\end{figure*}

\phantom{a}
\textbf{Subcase III:} $n=3$
\\
\\
The assumption $n=3$ means that we are dealing with $R^3$ gravity
with noncommutative geometry.  Now the shape function becomes

\begin{equation}
b(r) = m_0 \left[ -6r  \theta e^{-\frac{r^2}{12\theta}} + 6
  \theta^{\frac{3}{2}} (3\pi)^{\frac{1}{2}} erf \left
\{{\frac{r}{2\sqrt{3\theta}}}\right\}+C \right],
\end{equation}
where   $$m_0 = \Big(\frac{M}{4  a (4\pi \theta)^{\frac{3}{2}}
}\Big)^{\frac{1}{3}}.$$

\begin{figure*}[thbp]
\begin{tabular}{rl}
\includegraphics[width=4.5cm]{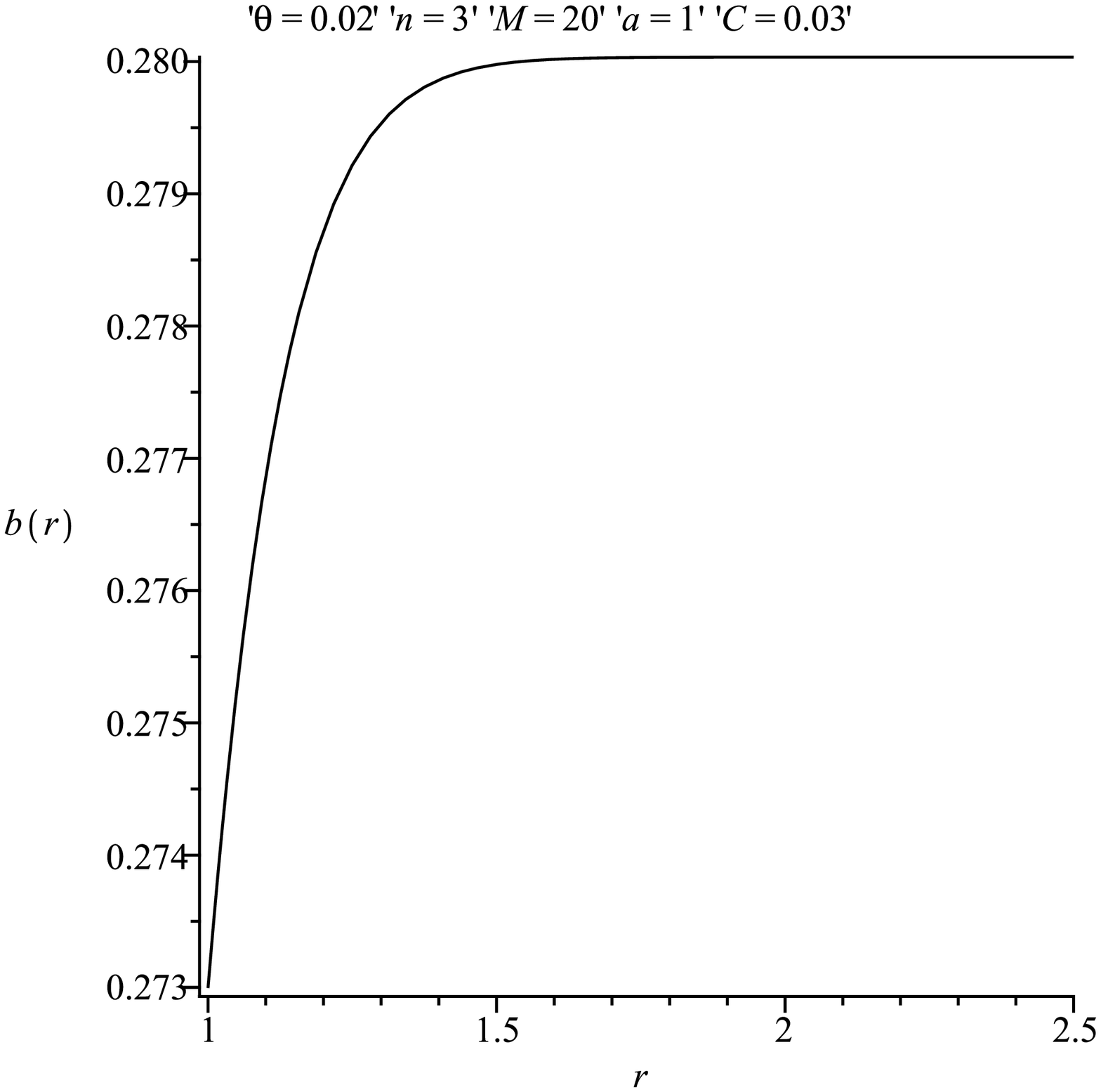}&
\includegraphics[width=4.5cm]{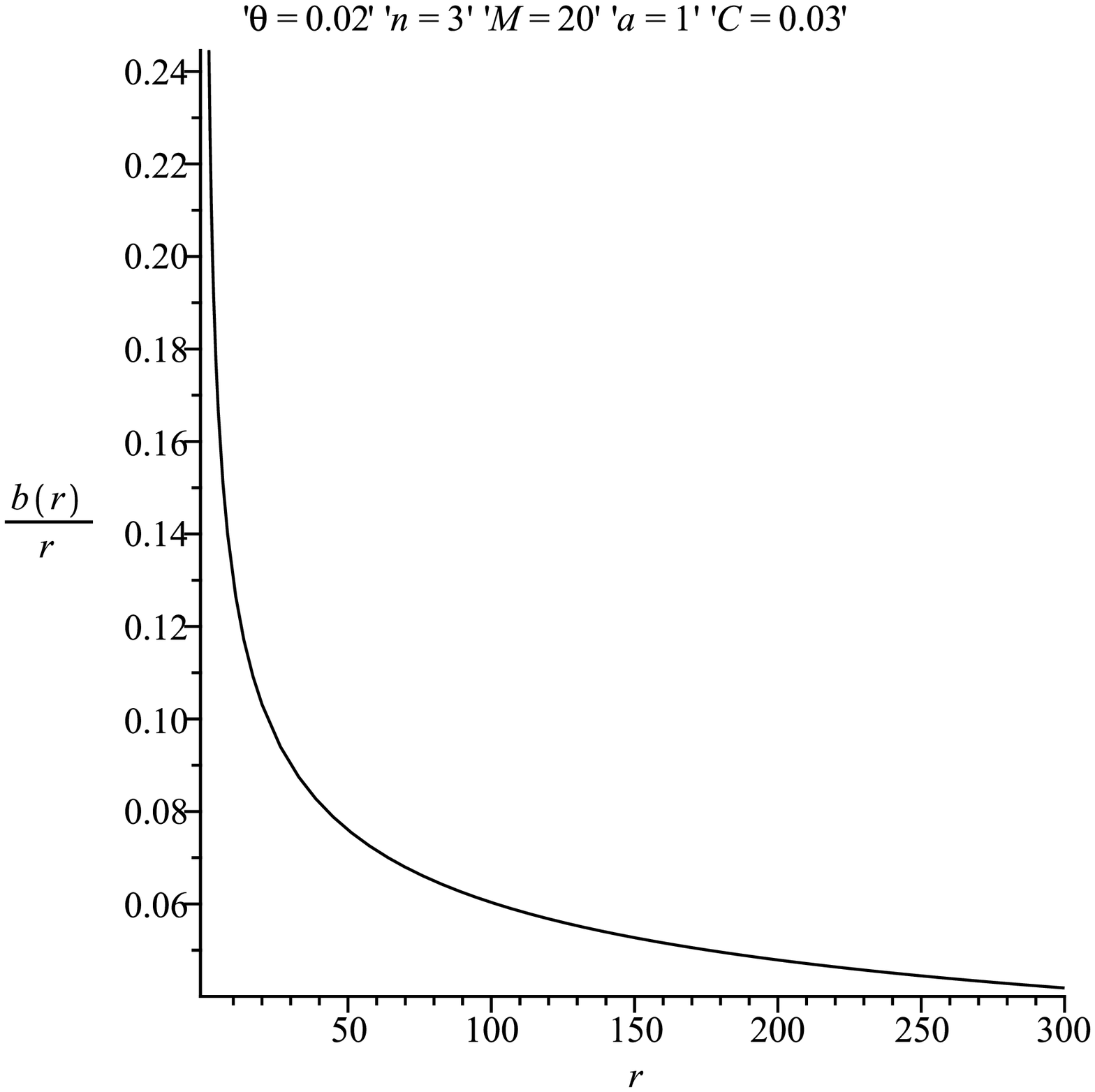}
\includegraphics[width=4.5cm]{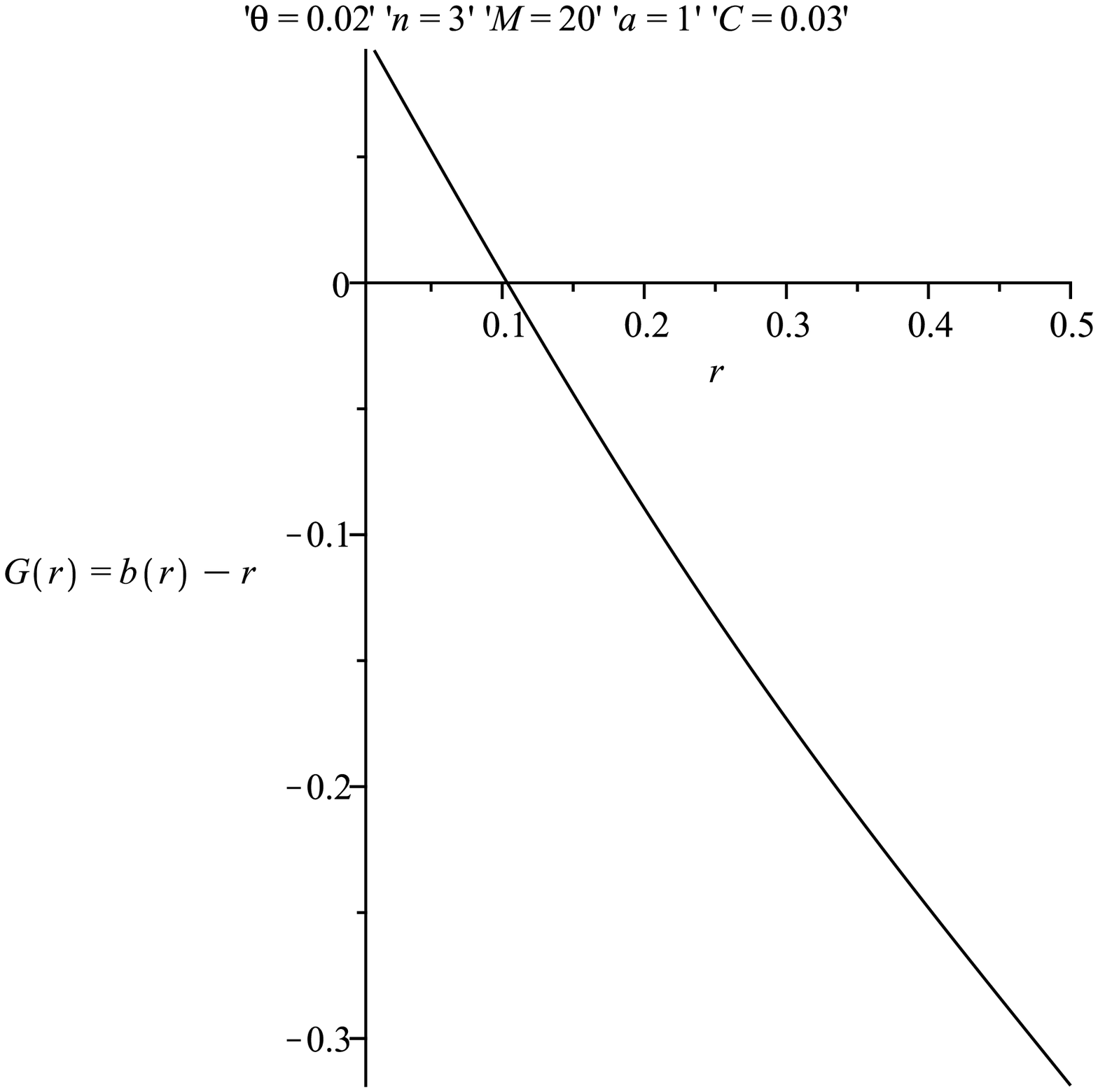} \\
\end{tabular}
\caption{ (\textit{Left}) Diagram of the shape function of
the wormhole in $R^3$ gravity for specific values of the
parameters: $\theta =0.02$, $M=20$, $a=1$, and $C=0.03$.
(\textit{Middle})  Asymptotic behavior of the shape function.
(\textit{Right}) The throat occurs where $G(r) = b(r)-r$
cuts the $r$-axis.}
\end{figure*}

The radial pressure component is given by

\begin{eqnarray}
p_r(r)= -\frac{4 a m_0^2e^{-\frac{r^2}{12\theta}}}{r^3}\left[m_0
\left\{ -6r  \theta e^{-\frac{r^2}{12\theta}} + 6
  \theta^{\frac{3}{2}} (3\pi)^{\frac{1}{2}} erf \left
({\frac{r}{2\sqrt{3\theta}}}\right)+C \right\} \right]
\nonumber\\-\frac{4 a m_0^2e^{-\frac{r^2}{6\theta}}}{6 \theta r
}\left[ m_0 r^3e^{-\frac{r^2}{12\theta}} -m_0 \left\{ -6r  \theta
e^{-\frac{r^2}{12\theta}} + 6
  \theta^{\frac{3}{2}} (3\pi)^{\frac{1}{2}} erf \left
({\frac{r}{2\sqrt{3\theta}}}\right)+C \right\} \right] \nonumber\\
-\left[ -\frac{4 a m_0^2e^{-\frac{r^2}{6\theta}}}{3
\theta}+\frac{4 a m_0^2 r^2 e^{-\frac{r^2}{6\theta}}}{9 \theta^2}
\right] \left[1 -  \frac{m_0}{r} \left\{ -6r \theta
e^{-\frac{r^2}{12\theta}} + 6
  \theta^{\frac{3}{2}} (3\pi)^{\frac{1}{2}} erf \left
({\frac{r}{2\sqrt{3\theta}}}\right)+C \right\}\right],
\end{eqnarray}

while the transverse pressure component is

\begin{eqnarray}
p_t(r)=\left[ \frac{4 a m_0^2e^{-\frac{r^2}{6\theta}}}{3 \theta}
\right]    \left[1 -  \frac{m_0}{r} \left\{ -6r \theta
e^{-\frac{r^2}{12\theta}} + 6
  \theta^{\frac{3}{2}} (3\pi)^{\frac{1}{2}} erf \left
({\frac{r}{2\sqrt{3\theta}}}\right) +C\right\}\right]
\nonumber\\
-\frac{2 a m_0^2e^{-\frac{r^2}{6\theta}}}{ r^3 }\left[ m_0
r^3e^{-\frac{r^2}{12\theta}} -m_0 \left\{ -6r  \theta
e^{-\frac{r^2}{12\theta}} + 6
  \theta^{\frac{3}{2}} (3\pi)^{\frac{1}{2}} erf \left
({\frac{r}{2\sqrt{3\theta}}}\right) +C\right\} \right].
\end{eqnarray}

\begin{figure*}[thbp]
\begin{tabular}{rl}
\includegraphics[width=5.5cm]{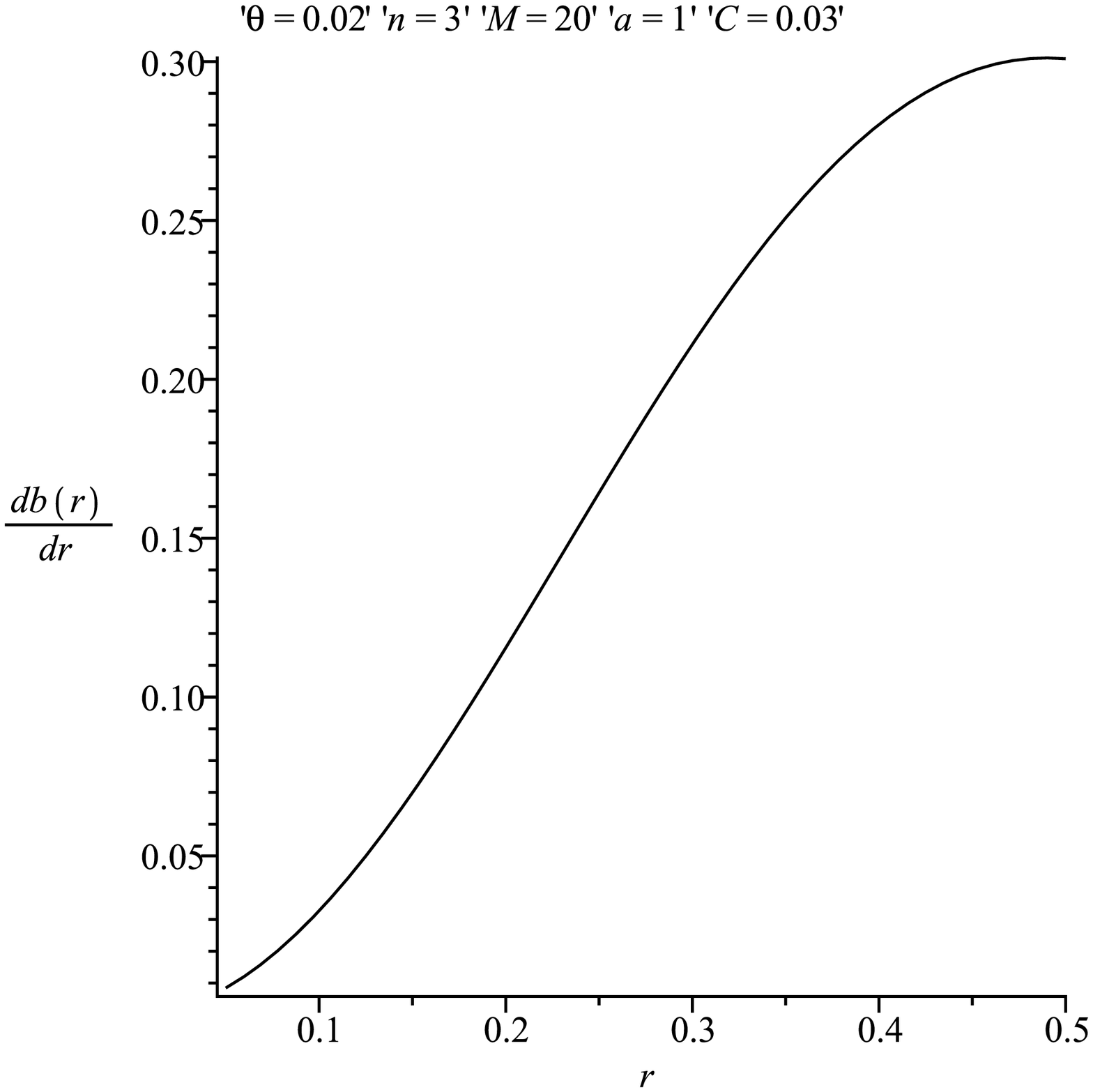}&
\includegraphics[width=5.5cm]{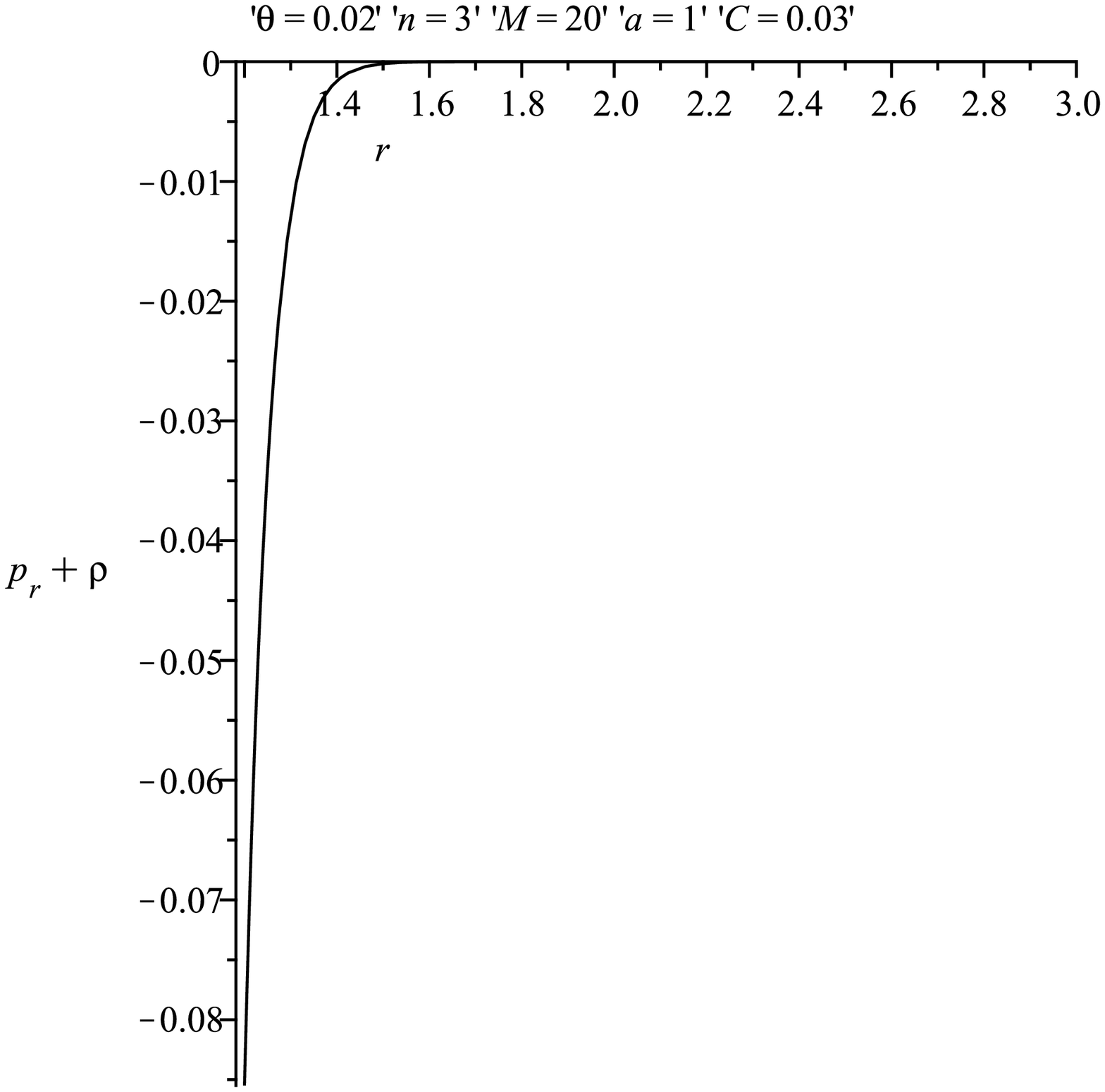} \\
\end{tabular}
\caption{ (\textit{Left})     Diagram of the derivative of
the shape function of the wormhole. (\textit{Right}) The
variations of the left-hand side of the expression for the null
energy condition with respect to $r$.}
\end{figure*}

Fig. 3 illustrates all the necessary characteristics of the shape function
of wormhole, while Fig. 4, in addition to the derivative of the shape
function, shows that the null energy condition is violated.
\\
\\
\textbf{Subcase IV:} $n=4$
\\
\\
The assumption $n=4$ yields $R^4$ gravity  with
noncommutative geometry.  Here the shape function, radial pressure,
and transverse pressure take on the respective forms

\begin{equation}
b(r) = m_0 \left[ -8r  \theta e^{-\frac{r^2}{16\theta}} +8
  \theta^{\frac{3}{2}} (4 \pi)^{\frac{1}{2}} erf \left
\{{\frac{r}{2\sqrt{ 4 \theta}}}\right\} +C\right],
\end{equation}
where   $$m_0 = \Big(\frac{M}{8  a (4\pi \theta)^{\frac{3}{2}}
}\Big)^{\frac{1}{4}}$$
\begin{eqnarray}
p_r = -\frac{8 a m_0^2e^{-\frac{3r^2}{16\theta}}}{r^3}\left[
m_0\left\{ -8r \theta e^{-\frac{r^2}{16\theta}} +8
  \theta^{\frac{3}{2}} (4 \pi)^{\frac{1}{2}} erf \left
({\frac{r}{2\sqrt{ 4 \theta}}}\right)+C \right\}\right]
\nonumber\\
-\frac{3 a m_0^3e^{-\frac{3r^2}{16\theta }}}{2 \theta r
}\left[ m_0 r^3e^{-\frac{r^2}{16\theta}} -m_0\left\{ -8r \theta
e^{-\frac{r^2}{16\theta}} +8
  \theta^{\frac{3}{2}} (4 \pi)^{\frac{1}{2}} erf \left
({\frac{r}{2\sqrt{ 4 \theta}}}\right) +C\right\} \right]
 \nonumber\\
-\left[ -\frac{3 a m_0^3e^{-\frac{3r^2}{16\theta}}}{
\theta}+\frac{9 a m_0^3 r^2 e^{-\frac{3r^2}{16\theta}}}{8
\theta^2} \right] \left[1 -  \frac{m_0}{r} \left\{ -8r \theta
e^{-\frac{r^2}{16\theta}} +8
  \theta^{\frac{3}{2}} (4 \pi)^{\frac{1}{2}} erf \left
({\frac{r}{2\sqrt{ 4 \theta}}}\right)+C   \right\}\right],
\end{eqnarray}
\begin{eqnarray}
p_t(r)=\left[ \frac{3 a m_0^3e^{-\frac{3r^2}{16\theta}}}{  \theta}
\right]    \left[1 -  \frac{m_0}{r} \left\{ -8r \theta
e^{-\frac{r^2}{16\theta}} +8
  \theta^{\frac{3}{2}} (4 \pi)^{\frac{1}{2}} erf \left
({\frac{r}{2\sqrt{ 4 \theta}}}\right)  +C \right\}\right]
\nonumber\\
-\frac{4 a m_0^3e^{-\frac{3r^2}{16\theta}}}{ r^3 }\left[ m_0
r^3e^{-\frac{r^2}{16\theta}} -m_0  \left\{ -8r \theta
e^{-\frac{r^2}{16\theta}} +8
  \theta^{\frac{3}{2}} (4 \pi)^{\frac{1}{2}} erf \left
({\frac{r}{2\sqrt{ 4 \theta}}}\right)  +C \right\} \right].
\end{eqnarray}

\begin{figure*}[thbp]
\begin{tabular}{rl}
\includegraphics[width=4.5cm]{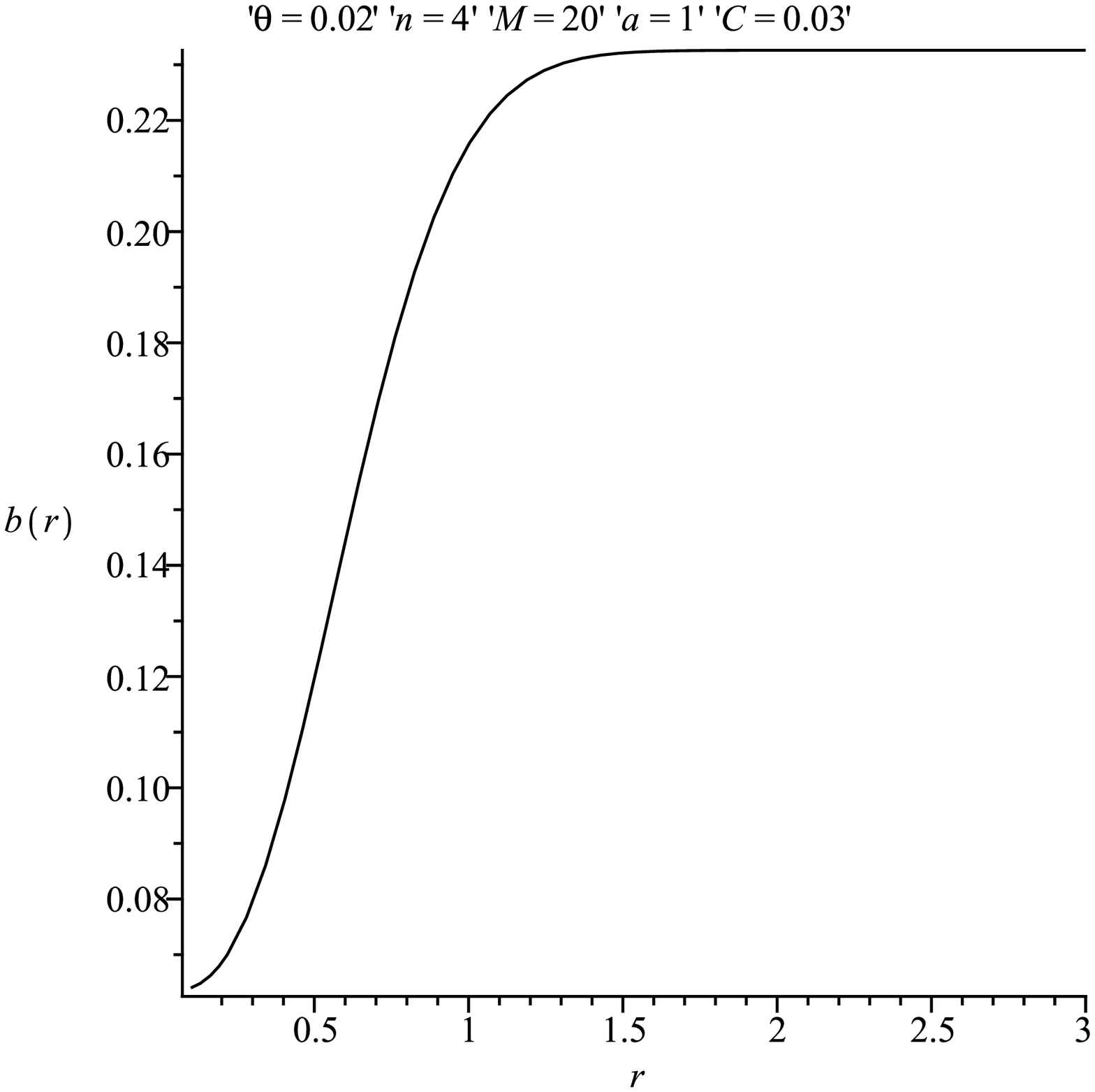}&
\includegraphics[width=4.5cm]{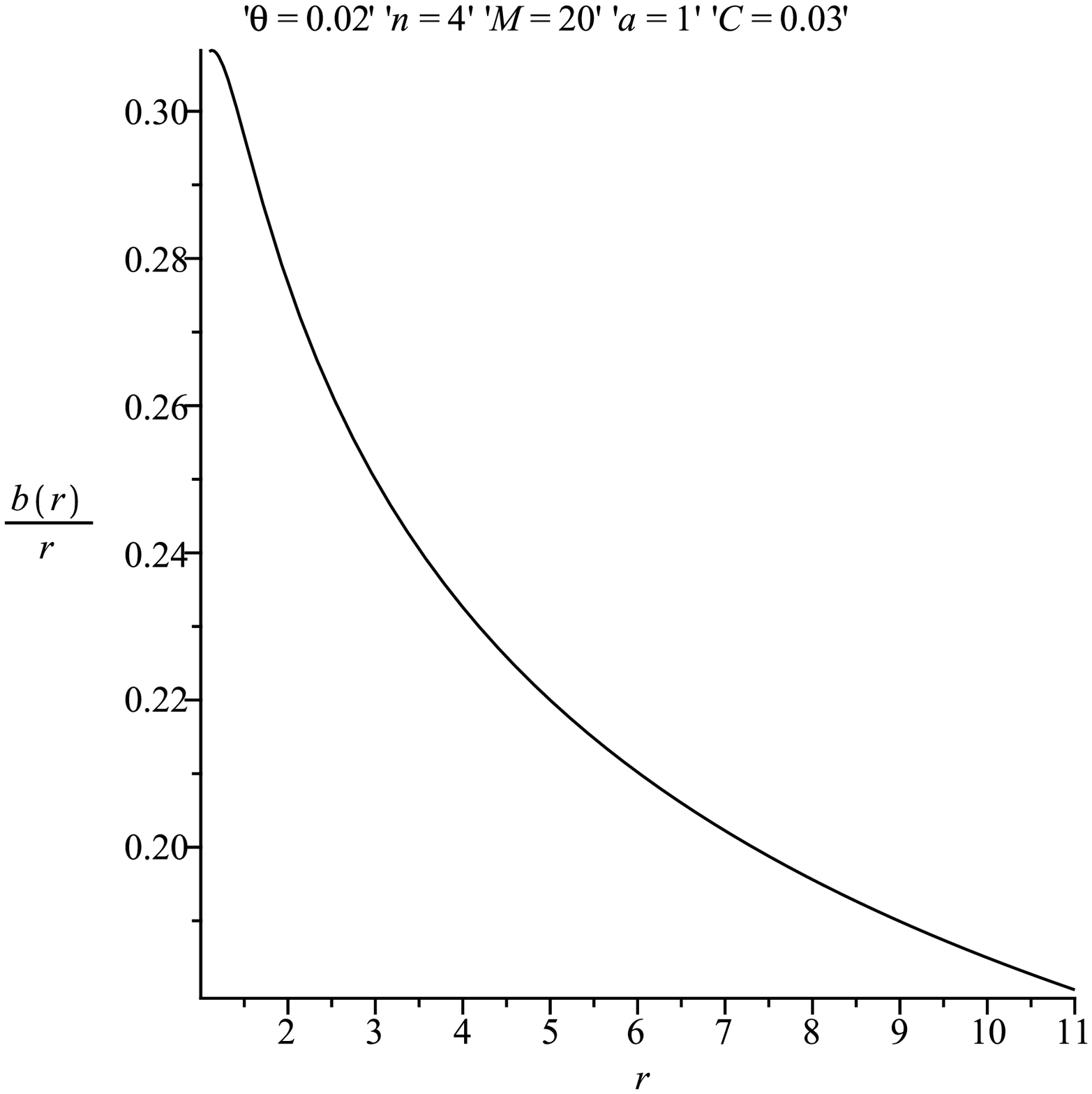}
\includegraphics[width=4.5cm]{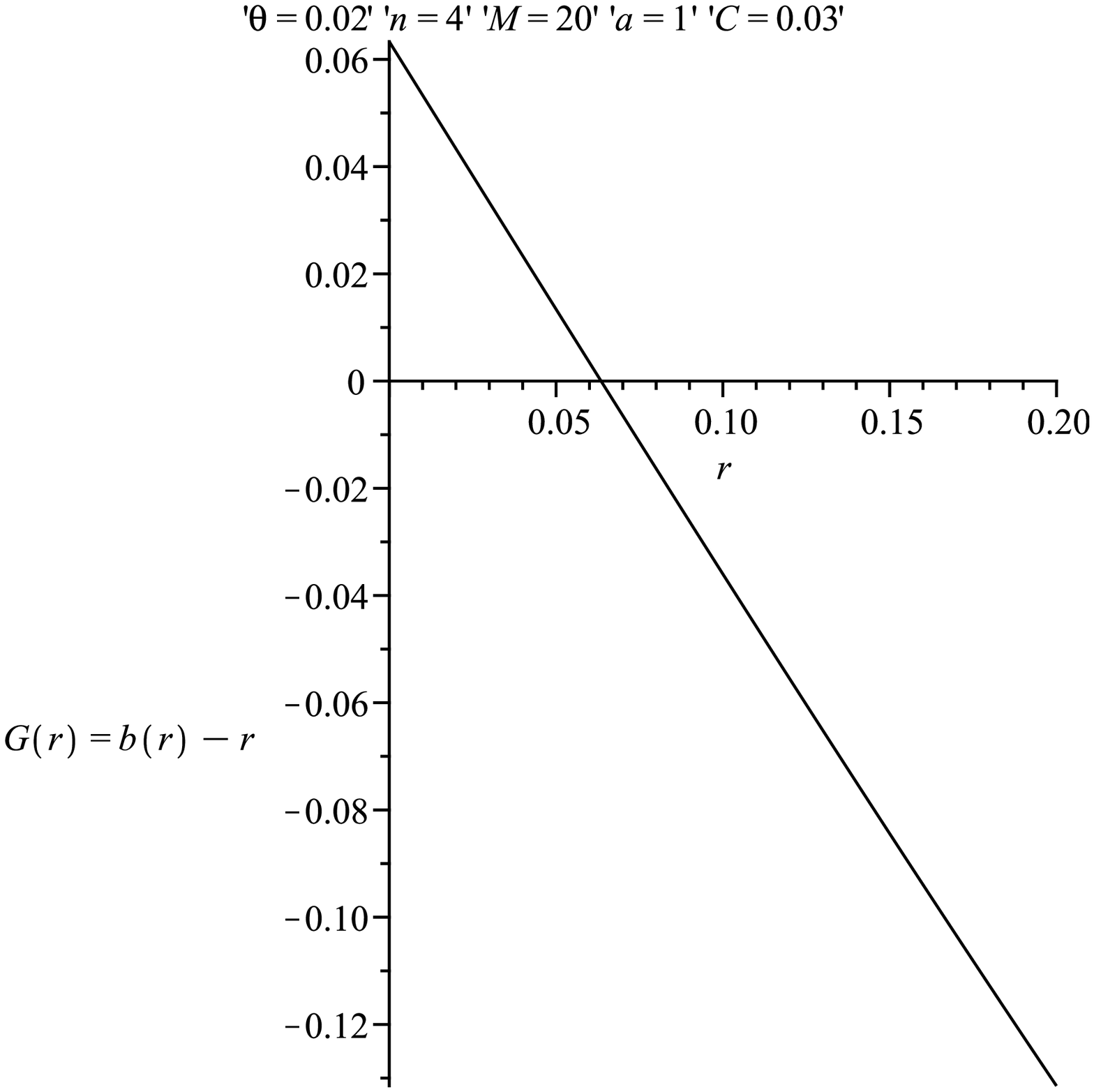} \\
\end{tabular}
\caption{ (\textit{Left})  Diagram of the shape function of
the wormhole in $R^4$ gravity for specific values of the
parameters: $\theta =0.02$, $M=20$, $a=1$, and $C=0.03$.
(\textit{Middle}) Asymptotic behavior of the shape function.
(\textit{Right}) The throat occurs where $G(r) = b(r)-r$
cuts the $r$-axis.}
\end{figure*}

\begin{figure*}[thbp]
\begin{tabular}{rl}
\includegraphics[width=5.5cm]{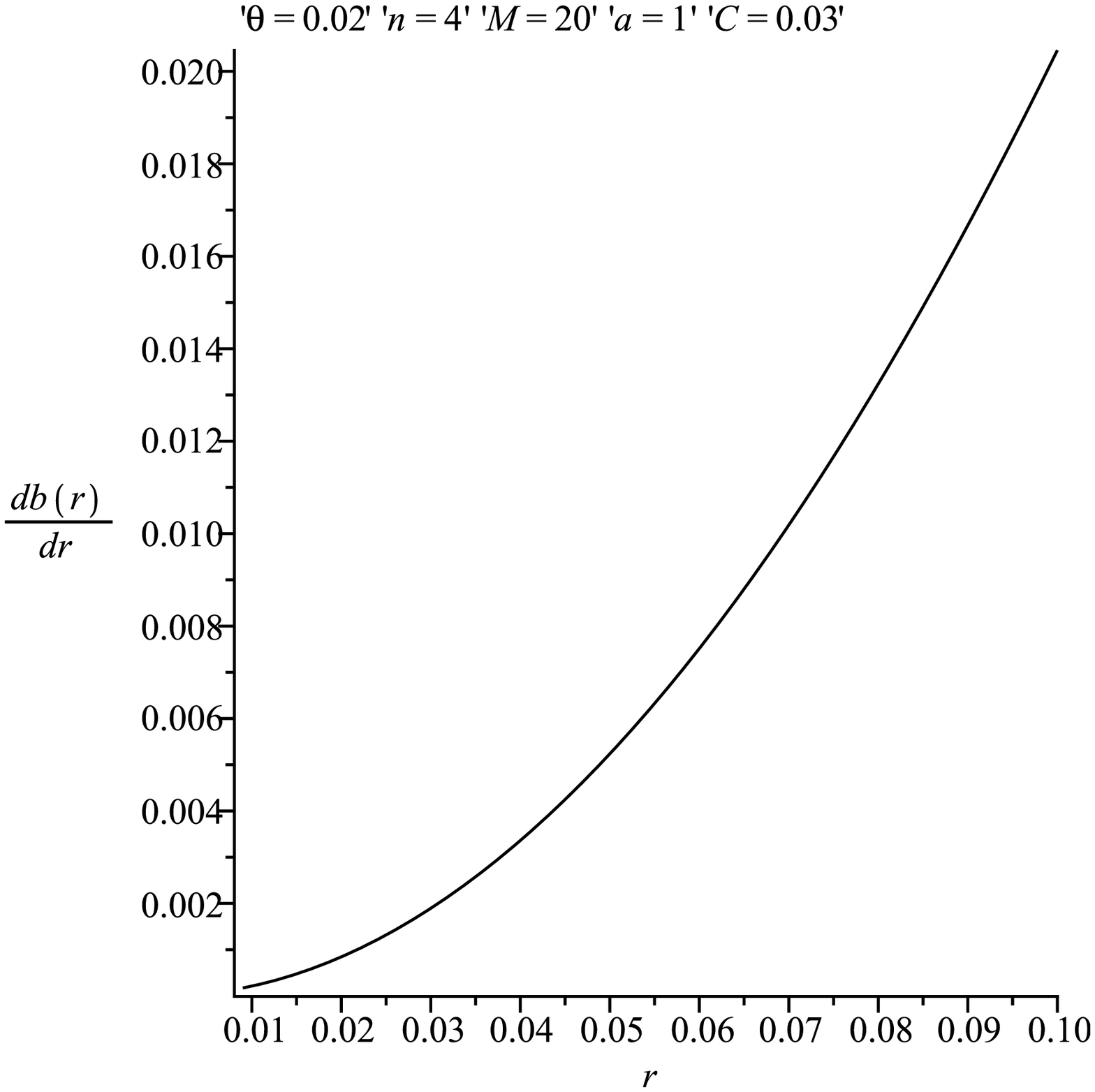}&
\includegraphics[width=5.5cm]{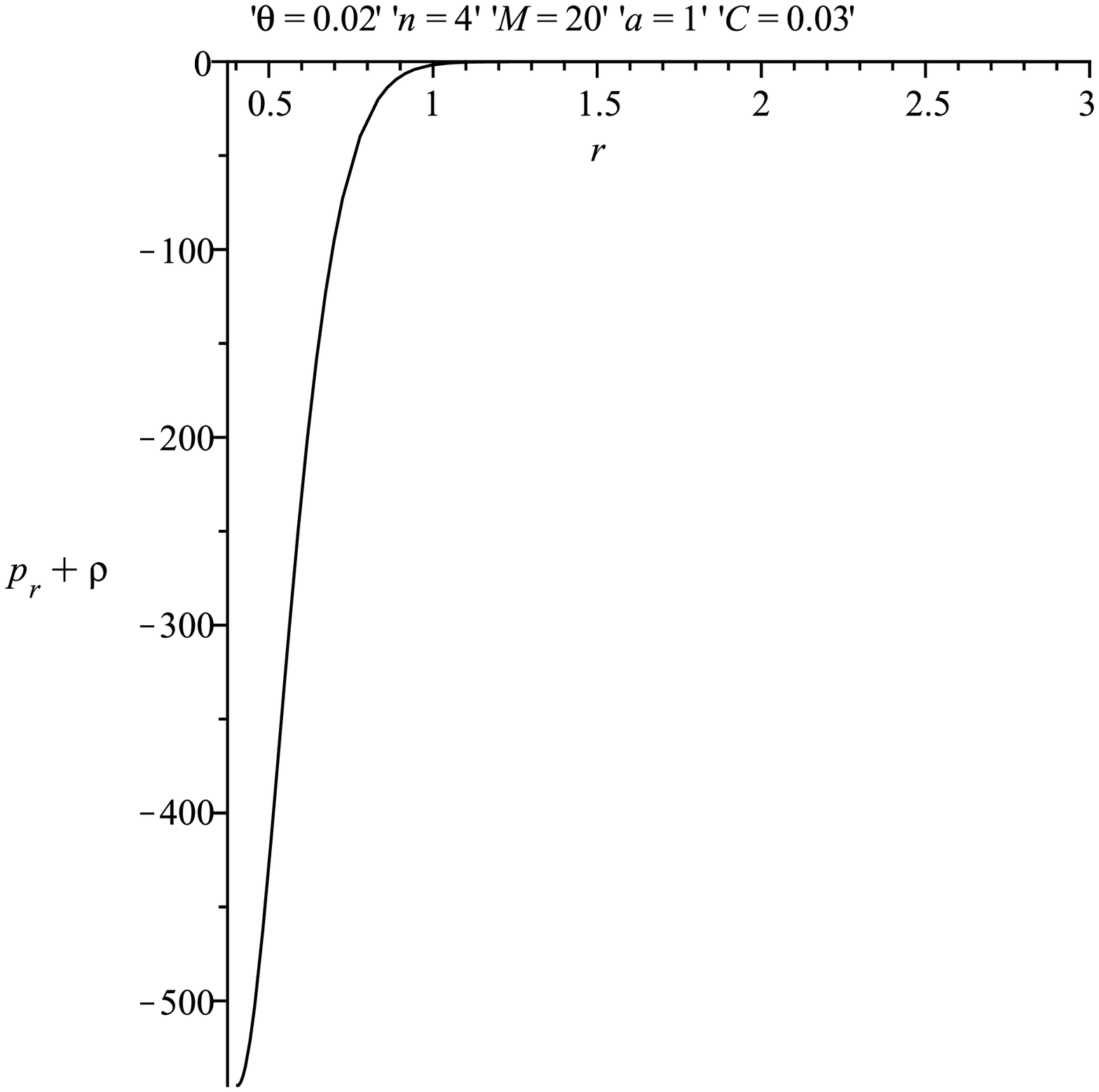} \\
\end{tabular}
\caption{ (\textit{Left})    Diagram of the derivative of
the shape function of the wormhole. (\textit{Right}) The
variation of the left-hand side of the expression for the null
energy condition with respect to $r$.}
\end{figure*}

As before, Figs. 5 and 6 illustrate the various characteristics
of the wormhole.

\phantom{a}

\textbf{Summary of Section III:} The power-law form of the
modified gravity, $F(R)=aR^{n-1}$, with a noncommutative-geometry
background yielded wormhole solutions with very similar
characteristics, including the violation of the null energy
condition.  The most striking feature is a decreasing throat size
as the power of $F(R)$ increases.

\section{Wormhole solution for a given shape function}
\noindent
This section discusses another wormhole solution using a
particular shape function \cite{ji}.  For this case, the null
energy condition is violated but the strong energy condition is
met.  The function is
\begin{equation}\label{br}
b(r)=r_0\Big(\frac{r}{r_0}\Big)^\alpha,\quad \alpha<1.
\end{equation}
Observe that $b(r_0)=r_0$ and $b'(r)=\alpha<1$, so that the
flare-out condition is satisfied.  From  Eq. (\ref{r}) we
obtain
\begin{equation}\label{fr1}
F(r)=\frac{\rho r^2}{b'(r)}=\frac{\rho r^{3-\alpha}}{\alpha
r_0^{1-\alpha}}.
\end{equation}
Moreover, the curvature scalar is given by
\begin{equation}
R=2\alpha\frac{r^{\alpha-3}}{r_0^{\alpha-1}}<\infty.
\end{equation}
So the curvature scalar is finite since $r_0>0$.  Using the
above expressions, it is easy to reconstruct $F(R)$:
\begin{equation}
F(R)=\frac{R}{2\alpha^2}\exp\Big[ -\frac{1}{4\theta} \Big\{
\frac{R}{2\alpha}\Big(\frac{2\alpha}{R_0}\Big)^{2(1-\alpha)}
\Big\}^{\frac{2}{3-\alpha}} \Big];
\end{equation}
here $R_0$ is the value of curvature scalar at $r=r_0$.

Making use of Eqs. (\ref{br}) and (\ref{fr1}) in (\ref{r1})
and (\ref{r2}), we get the radial pressure

\begin{eqnarray}
p_r&=&
\frac{M}{(4\pi\theta)^{\frac{3}{2}}}e^{-\frac{r^2}{4\theta}} \Big[
-\frac{r^\alpha}{3\alpha}+\frac{r^{4-2\alpha}}{r_0^{2(1-\alpha)}4\alpha^2(\alpha-3)}\Big(
-\frac{r^2}{2\theta}+3-\alpha \Big)
r_0^{1-\alpha}r^\alpha(\alpha-1)\nonumber\\&&-
\frac{r^{9-3\alpha}}{2\alpha(\alpha-3)r_0^{1-\alpha}}\Big\{
-\frac{r^2}{4\theta r_0^{2(1-\alpha)}\alpha^2(\alpha-3)}\Big(
-\frac{r^2}{2\theta}+3-\alpha\Big)  \nonumber\\&&
+\frac{1}{r_0^2(\alpha-1)}\Big(
-\frac{r^2}{2\theta}+3-\alpha\Big)+\frac{r}{r_0^{2(1-\alpha)}2\alpha^2(\alpha-3)}\Big(
-\frac{r}{\theta}+3-\alpha\Big)\Big\}(1-r_0^{1-\alpha}r^{\alpha-1})
\Big]
\end{eqnarray}
as well as the transverse pressure
\begin{eqnarray}
p_t(r)&=&\frac{M}{(4\pi\theta)^{\frac{3}{2}}}e^{-\frac{r^2}{4\theta}}\Big[
-\frac{r^{5-2\alpha}}{r_0^{2(1-\alpha)2\alpha^2(\alpha-3)}} \Big(
-\frac{r^2}{2\theta}+3-\alpha
\Big)(1-r_0^{1-\alpha}r^{\alpha-1})\nonumber\\&&+\frac{r^{3-\alpha}}{r_0^{1-\alpha}4\alpha^2(\alpha-3)}\Big(
-\frac{r^2}{2\theta}+3-\alpha \Big)(1-\alpha) \Big].
\end{eqnarray}
The physical viability of this solution is determined by checking
the energy conditions.  We can see from Fig. 7 that $p_r+\rho<0$.
It is interesting to note the strong energy condition is satisfied.

\begin{figure*}[thbp]
\begin{tabular}{rl}
\includegraphics[width=7.5cm]{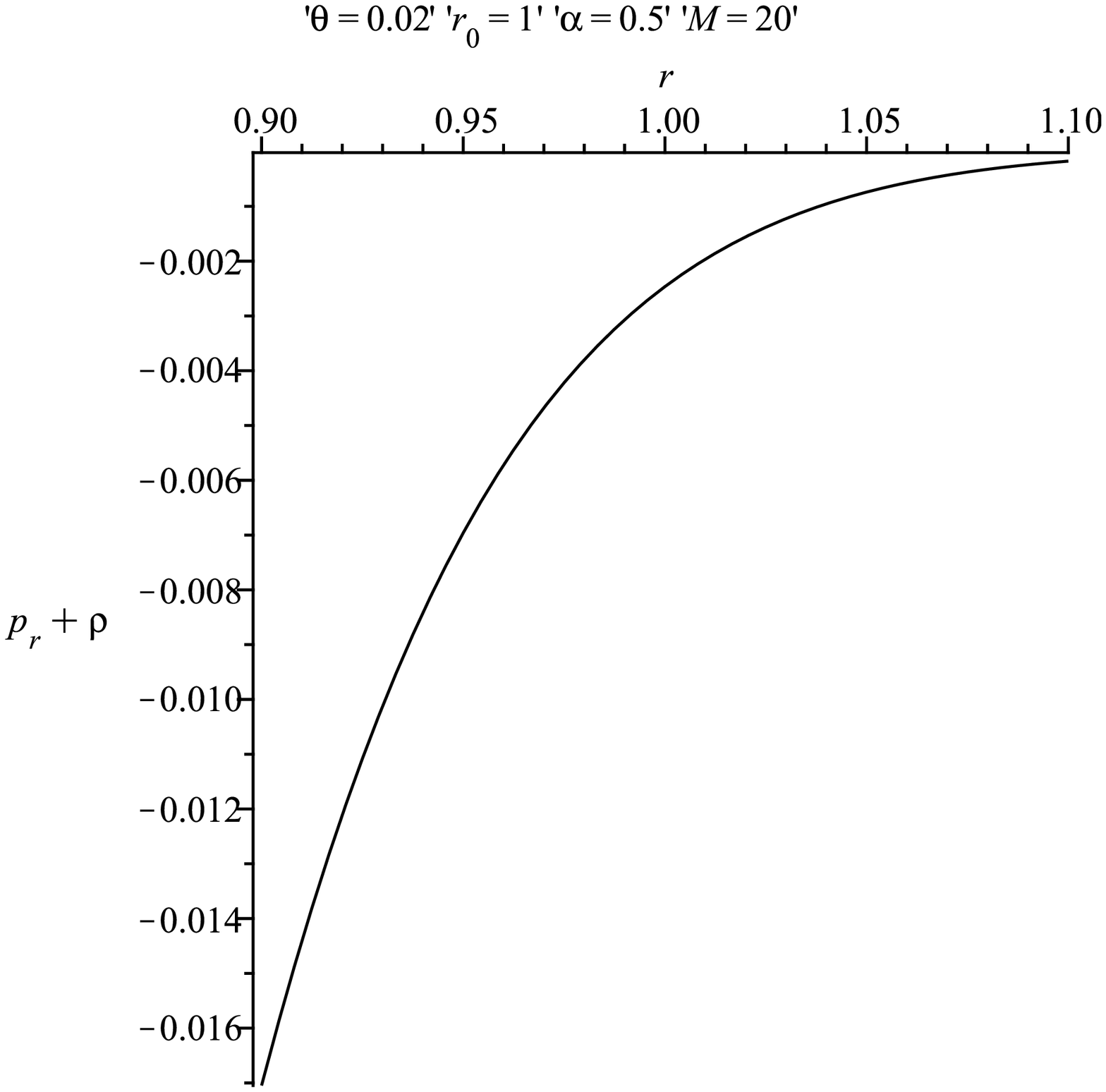}&
\includegraphics[width=7.5cm]{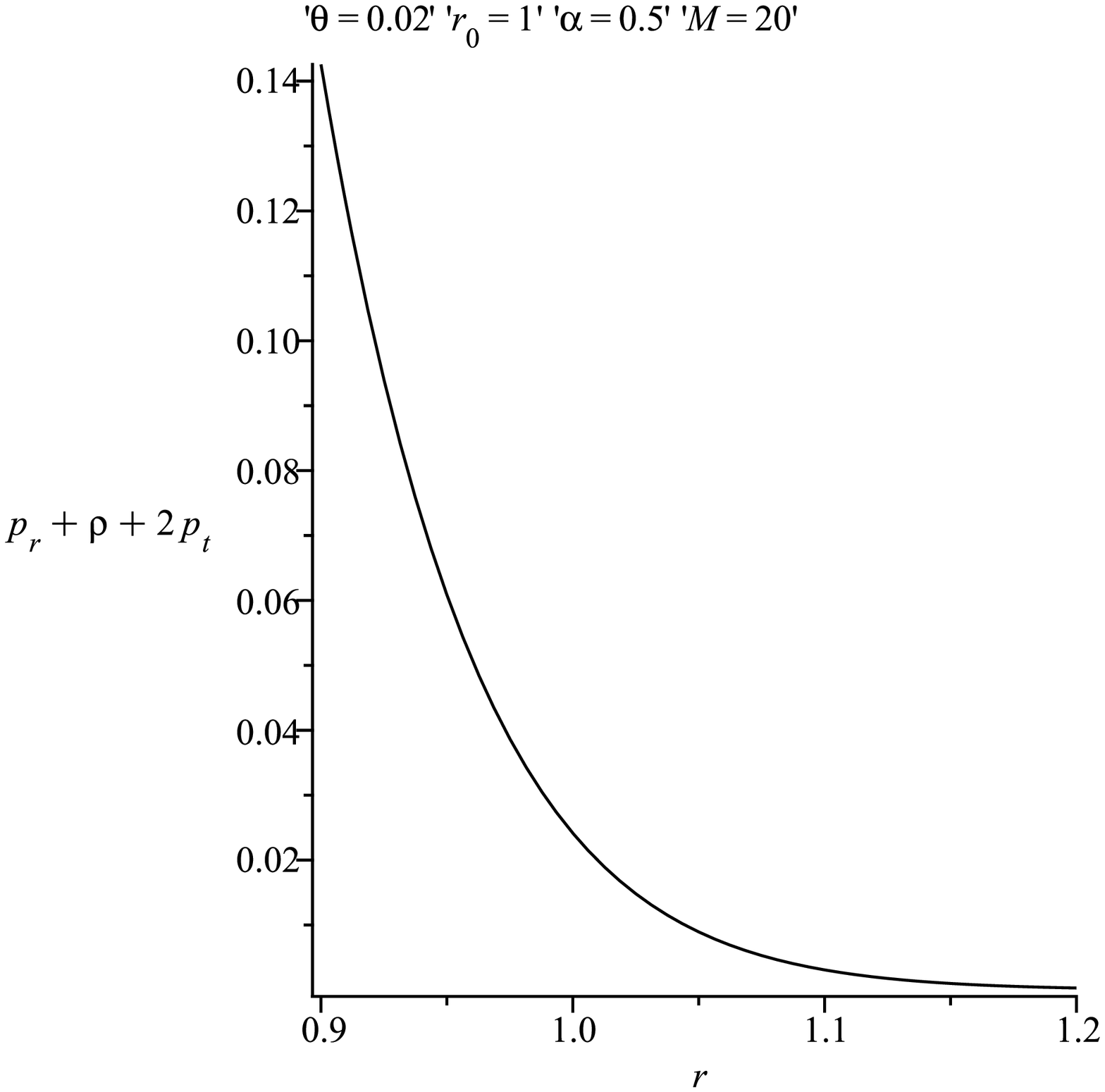} \\
\end{tabular}
\caption{ (\textit{Left})  The variation of the radial null energy
conditions with respect to $r$. (\textit{Right}) The behavior of the
strong energy condition is shown against $r$.}
\end{figure*}

\section{Conclusion}\noindent
Noncommutative geometry, an offshoot of string theory, replaces
pointlike structures by smeared objects and has recently been
extended to higher dimensions. In this paper we present two models
of wormholes within the framework of this extended noncommutative
geometry. The first model assumes the power-law form
$F(R)=aR^{n-1}$. The analysis includes the important special cases
$n=1$ (Einstein gravity), already discussed in Ref. \cite{farook},
and $n=2$ ($R^2$ gravity). It is shown that the basic
characteristics, particularly the violation of the null energy
condition, remain essentially the same, but the radius of the
throat decreases as the power of $F(R)$ increases.  The second
model discussed assumes a particular shape function, thereby
allowing a reconstruction of $F(R)$.  For this case, the null
energy condition is once again violated, but the strong energy
condition is met.  All the solutions assume zero tidal forces, a
highly desirable feature from the standpoint of wormhole design.

\section*{Acknowledgments}
\noindent FR and NA wish to thank the authorities of the
Inter-University Centre for Astronomy and Astrophysics, Pune, India,
for providing them Visiting Associateships under which a part of
this work was carried out. FR also thanks  UGC, Govt. of India, for
providing financial support under the research support scheme. MJ
has been supported from the grant of Higher Education Commission,
Islamabad, project No. 20-2166.

\end{document}